\documentstyle[prl,aps]{revtex}
\draft
\input epsf

\begin{document}

\title{Coiling Instability of Multilamellar Membrane Tubes \\
with Anchored Polymers}
\author{ Ilan Tsafrir$^1$, Marie-Alice 
Guedeau-Boudeville,$^2$ \\
Daniel Kandel$^{1}$, and Joel Stavans$^1$}
\address{$^1$ Department of Physics of Complex Systems, \\
The Weizmann Institute of Science,
Rehovot 76~100, Israel \\
      \vspace*{3 mm}
$^2$ Laboratoire de Physique de la Mati\`{e}re Condens\'{e}e, URA 792,
Coll\`{e}ge de France\\
11 Place Marcelin Berthelot,
F-75231 Paris CEDEX 05, France \\
      \vspace*{3 mm}
}
\maketitle

\begin{abstract}
We study experimentally a coiling instability of cylindrical 
multilamellar stacks of phospholipid membranes, induced by polymers 
with hydrophobic anchors grafted along their hydrophilic backbone.  
Our system is unique in that coils form in the absence of both twist 
and adhesion.  We interpret our experimental results in terms of a 
model in which local membrane curvature and polymer concentration are 
coupled.  The model predicts the occurrence of maximally tight coils 
above a threshold polymer occupancy.  A proper comparison between the 
model and experiment involved imaging of projections from simulated 
coiled tubes with maximal curvature and complicated torsions.

\end{abstract}

\vspace{0.3cm}
{\hspace*{2.cm}}PACS numbers: 87.16.Dg, 68.10.-m
\vspace{0.3cm}

\section{Introduction}

Coiling is a common occurrence in vastly different systems and on many 
scales, ranging from carbon nano-tubes \cite{Zhong-can:97} and DNA 
molecules \cite{Cozzarelli:90}, to telephone cords and tendrils 
of climbing plants \cite{Goriely:98}.  The framework most commonly 
used to model coiling in these and other systems is that of an elastic 
rod.  Buckling of the central line is explained by showing that 
converting twist to writhe lowers the elastic energy 
\cite{klapper:96}.


We study the coiling of cylindrical stacks of lipid membranes 
\cite{frette:99}, called 
\emph{myelin figures}, interacting with an amphiphilic polymer.  This 
polymer has several hydrophobic side groups attached along a flexible 
hydrophilic backbone, which it inserts into the membranes in order to 
shield them from the surrounding water \cite{Ringsdorf:91}.  The 
membranes composing a myelin figure are in a two-dimensional liquid 
state, and therefore cannot support twist.  Application of torque on 
the cylinder simply leads to flow of material around the tube.  Thus 
the interplay between twist and writhe cannot explain the coiling 
observed in our experiments.

The existence and coiling of myelin figures has been observed as far 
back as 1854 \cite{virchow:54} (see also \cite{Lehmann:13}).  More
recently, similar shapes have 
been observed during the hydration of a surfactant by brine 
\cite{Buchanan:98,Buchanan:00}.  Coiling of myelin figures has also been
reported 
for a system of egg-yolk lecithin 
\cite{Sakurai:85,Sakurai:89,Mishima:92}.  This system is a mixture 
containing many different lipids, having a variety of tail lengths and 
degrees of saturation \cite{Tattrie:68}.  In another experiment it was 
shown that a binary mixture of dimyristoylphosphatidylcholine and 
cardiolipin forms single and double helices in the presence of calcium 
ions \cite{Lin:82}.  In these two studies, it was claimed that: 
\emph{(i)} the energy gained by surface adhesion overcomes the energy 
cost of bending a tube, and \emph{(ii)} the tighter the coil, the 
longer the line of contact between tubes becomes, thus increasing the 
area of contact.

In contrast, our experiments clearly show that surface adhesion is 
negligible in our system. 
In order to account for coiling in our experiments, we present a 
simple model, in which we assume that polymer molecules locally induce 
spontaneous curvature.  The coiling instability results from a 
coupling between local polymer concentration and membrane curvature.  
Within such a framework, hollow tubes may also undergo other shape
instabilities such as pearling
\cite{Ringsdorf:91,Tsafrir:00,Deuling:77}.  
However, the constraints imposed by the geometry of the cylindrical 
membrane stack prevent them from occurring.  
Our model predicts the occurrence of maximally tight coils above a 
threshold polymer concentration.  Indeed, only maximally curved 
coils were formed in our experiments. A theoretical analysis 
shows that virtual slices through maximally curved helices can be very 
similar to the observed images.

Section \ref{Materials} describes the materials and methods used in 
our experiments.  Section \ref{results} describes the types of coils 
observed, and shows that adhesion does not play a major role in these 
experiments.  Section \ref{theory} presents the simple model that 
explains the major findings.  First, a heuristic argument is given as 
to why such an approach works.  Then the full mean-field calculation 
is given.  Finally, section \ref{image-model} outlines the difficulties 
in comparing quantitatively the model and the experiment.

\section{Materials and Methods}
\label{Materials}
Vesicles were made of stearoyl-oleoyl-phosphatidylcholine (SOPC) with 
\(C_{18}\) alkyl chains.  The polymer used is hydrophilic dextran (MW 
162,000 g/mol) functionalized both with palmitoyl alkyl chains and 
dodecanoic NBD chains as fluorescent markers.  The hydrophobic 
anchors, distributed statistically along the backbone (about 1 alkyl 
chain per 25 glucose units) are \(C_{16}\) long.  On average there 
are about 4 persistence lengths between consecutive anchors.  
Therefore, the extension of each polymer molecule on the 
two-dimensional membrane is much larger than its extension into the 
third dimension.
Events were observed by phase contrast microscopy and recorded on 
video.  For fluorescence imaging the NBD markers were excited with 
argon laser illumination, and observed with a cooled CCD camera.  

Samples were prepared by drying a 0.5-1.0\,$\mu$l droplet of SOPC 
dissolved in a 4:1 chloroform-methanol solution (7.35\,mg/ml) on a 
glass slide.  The sample was then closed from the top and sides, and 
hydration was effected by injecting a polymer solution of known 
concentration, $c_p$, into the closed cell.  Some time after 
hydration, a variety of self-assembled structures formed, including 
myelin figures.  All the structures observed in these experiments were 
still connected to a lipid reservoir, allowing exchange of material, 
including both lipids and polymer molecules.  Experiments were 
conducted at room temperature, well above the solid-liquid transition 
for SOPC. This allowed free diffusion of anchored polymers along the 
membranes.

We stress that while the polymer concentration in solution, $c_p$, is 
known, we do not control the surface concentration on the bilayers.  
The slow evolution of some of the structures we observe is consistent 
with a possible variation of this concentration over time.

\section{results}
\label{results}


Electron micrographs of cross sections of myelin figures reveal that 
they are rodlike lyotropic liquid-crystalline structures containing a 
large number (hundreds to thousands) of concentric cylindrical 
membranes separated by thin hydration layers \cite{Sakurai:89}.  The 
smectic order in these stacks of membranes is not ideal, as many 
defects are present.  The outer radius of a myelin figure can reach 
tens of microns, while the radius of the water-core may be of the 
order of, or smaller than optical resolution (0.2$\mu$).  The myelin 
figures, which are connected at one end to a large lipid reservoir, 
are continuously elongating.  The rate of elongation of a myelin 
figure lies in the range of $0-0.3 \mu m/sec$.  Throughout the 
experiment, polymer molecules continue to anchor from the surrounding 
solution.  Hence we assume that there is continuous accumulation of 
both lipid and polymer molecules.

Hydration of a patch of lipids by a polymer solution of small $c_{p}$ 
results in the formation of myelin figures, which display a clear 
tendency to straighten over lengths many times larger than their 
diameter (see Fig.\  \ref{fig:straight}).  As \(c_{p}\) is increased, 
myelin figures become more floppy and curved.  For large enough values 
of \(c_{p}\), a writhing instability sets in and tubes bend, forming 
irregular structures (Fig.\  \ref{fig:ball}), single coils (Fig.\  
\ref{fig:spring}a) and double helices (Fig.\  \ref{fig:dbl-coil}).  The 
type of coiled structure formed, depends in great part on the dynamics 
of the formation process.
By far the most common event is for the tip to begin to curve in upon 
itself forming a seemingly irregular ball-like structure (Fig.\  
\ref{fig:ball}).  However, when the ball is large enough, some sort of 
ordering can be seen (Fig.\  \ref{fig:kink}e).
Coils that start forming at their bases usually evolve into nearly 
ideal single helices.  The uncoiled part of the tube leading to the 
tip is pulled in and wound around to form the next loop \cite{web-site}.
In cases where the coiling starts in the middle, it begins as a 
hairpin that rotates around itself, and the instability then proceeds 
in both directions (Fig.\  \ref{fig:kink}).  This may result either in 
the formation of a tightly packed sphere or a double helix.  At the 
site where the instability nucleates, the tube goes from nearly 
straight to maximally curved.  The instability then propagates from 
this site to the rest of the tube (Fig.\  \ref{fig:ball} and 
\ref{fig:kink}).  The observed evolution may be due to a gradual 
change in the concentration of polymer molecules on the membrane.

All the coiled structures we observe are maximally curved already 
\emph{as they form} and do not tighten up gradually,
unlike the experiments reported by Sakurai et 
al.  \cite{Sakurai:85,Sakurai:89}.  In quantitative terms, this means 
that the curvature of the tube central line, $C$, is 
$C\approx 1/r_0$, where $r_0$ is the radius of the tube.


In previously reported cases \cite{Sakurai:85,Lin:82}, coiling of 
myelin figures was attributed to surface adhesion.  This is clearly 
not the case in our system, as demonstrated by the experiment 
illustrated in Fig.\  \ref{fig:spring}.  The tip of a myelin figure, in 
the process of coiling, adhered to an air bubble, which we then moved.  
Movement of the bubble stretched the coil (Fig.\  \ref{fig:spring}a-c), 
until the latter reached a configuration in which all self-contact was 
lost (Fig.\  \ref{fig:spring}c). Note also that the coil is stretched 
more or less homogeneously.  Upon detachment from the bubble, the coil 
retracted as if it was an ordinary spring (Fig.\  \ref{fig:spring}d-g).

Had surface adhesion been the dominant mechanism, one would not expect 
the response of a coil to mechanical stretching to be homogeneous, but 
rather to come apart at the site of weakest contact. Furthermore, 
adhesion cannot create a restoring force. Thus, if the coil is 
stretched open so that no contact sites are left, the tube should 
``forget'' that it was coiled. If the force exerted on the end of 
the tube is then released, adhesion would induce coil formation starting
at one point and propagating to the rest of the tube (similarly to the 
original formation process). This is in stark contrast with our
experimental observations.

Another piece of evidence against adhesion in our system is provided by 
the presence of many other structures in our experiments that 
come into contact with one another, but do not adhere (see e.g. Fig.\  
\ref{fig:contact}).

In order to understand better the role that the polymer plays in the 
coiling phenomenon, it is important to know its location in the 
membrane stack.  For this purpose, fluorescence imaging was used.  The 
results are shown in Fig.\  \ref{fig:flo-tube}.  As can be seen, the 
fluorescence intensity through the slice is concave.  This is what one 
expects for a homogeneous polymer distribution throughout the stack, 
since in that case, the intensity should be roughly proportional to 
the thickness of the tube in the microscope slice.  From this we infer 
that the polymer is present, in significant quantity, throughout the 
myelin figure.  Images of hollow tubes, where the polymer sits 
predominantly on the outer layers, have a convex fluorescence 
intensity profile through their cross sections.

\section{Theoretical Model}
\label{theory}

We now present a simple theoretical model that captures most of the 
key experimental observations.  For simplicity, we regard the system 
as if it were in equilibrium, effectively ignoring the slow evolution 
of the observed structures.  

Consider a stack of concentric cylindrical sheets.  We 
represent each bilayer as two square lattices (in the spirit of 
lattice-gas models), corresponding to the outer and inner monolayers.  
Each site represents a patch of membrane having area $a^{2}$, 
approximately the size of a polymer molecule performing a
two-dimensional random walk on the membrane.  Each lattice site has two 
degrees of freedom associated with it: the local mean curvature, $H$, 
and a binary occupation variable, $\sigma$, that takes the values one 
or zero when the site is occupied or unoccupied by a polymer molecule, 
respectively.  To estimate the area of a site we assume that the 
hydrophilic backbone is in a good solvent (water) under semi-dilute 
conditions.  This gives a radius of gyration, $a \approx 40 - 80$ nm.

The energy of the system is a sum of the curvature energies of the 
sites: $2\kappa H^2$ for a vacant site, and $2\kappa'(H-H_0)^2$ for a 
site occupied by a polymer.  $\kappa$ and $\kappa'$ are the local 
bending rigidities of a monolayer without and with an attached 
polymer, respectively, and $H_{0}$ is the spontaneous curvature 
induced by the polymer.  We assume that $H_{0} > 0$, since the 
addition of polymer tends to bend membranes into shapes with higher 
curvatures.  
For the purpose of this simple model, the exact molecular mechanism 
responsible for this is not important. Possible mechanisms include 
the entropic pressure of the polymer backbone, or the 
incommensurability between the anchors and lipids. By convention, the
curvatures of the inner and outer 
monolayers of the bilayer have opposite signs at the same position.

A crucial assumption is that polymer molecules can diffuse along the 
membrane, since the membrane is in a fluid state.  In addition, we 
assume that the fluctuations of each membrane in a stack are severely 
restricted by the presence of its neighbors.  As a result, there is a 
strong correlation between the transversal fluctuations throughout the 
tube.  The myelin figure can thus be regarded as a flexible rod, 
having a circular cross section everywhere along its axis.  The 
experimental pictures indeed display unchanging circular 
cross sections within experimental error.

Based on these assumptions, we developed a model that predicts that a 
high enough polymer concentration on the membrane can shift the 
equilibrium state, from a straight tube to a maximally curved one.
We first present a heuristic argument to show that if the spontaneous 
curvature, $H_0$, is large enough, the free energy of a bent tube may 
be lower than that of a straight one.  This approach may give a more 
intuitive understanding than the detailed calculations that follow.  

\subsection{Heuristic Arguments}

In order to find the equilibrium state of a tube, we have to calculate 
its free energy.  This free energy depends on the curvature of its 
central line, $C$, and on the polymer concentration.  

Consider one cylindrical bilayer of length $l$ and circular cross 
section of radius $r$, with the same average polymer concentration, 
$\rho_{0}$, on both sides.  $\rho_{0}$ is defined as the number of 
polymer molecules on a monolayer, $n$, divided by the total number of 
sites on a monolayer, i.e. $\rho_{0}=na^2/A$, where $A$ is the total 
area of the membrane segment.  Let us calculate the free energy cost 
of bending the bilayer into a portion of a coil with central line 
curvature $C$ in three steps.  First, we bend the tube while keeping 
the distribution of polymer around the tube homogeneous.  Next, we 
allow the polymer to diffuse from regions of lower curvature to 
regions of higher curvature (Fig.\  \ref{fig:scheme}), and finally, we 
consider the entropy of mixing.

The energy of a bent cylindrical bilayer with a {\it homogeneous} 
polymer distribution is $E_{hom}=E_{hom}^{+} + E_{hom}^{-}$, where 
$E_{hom}^{+}$ and $E_{hom}^{-}$ are the energies of the outer and 
inner monolayers, respectively.  According to our model

\begin{eqnarray}
E_{hom}^{\pm}(C) & = & 2\rho_{0}\kappa'\int
dA\left[H^{\pm}(C,\phi)-H_0\right]^2 + 2(1-\rho_{0})\kappa\int
dA\left[H^{\pm}(C,\phi)\right]^2~,
\label{eq:ehom}
\end{eqnarray}

\noindent where $\phi$ is the angle around the tube, \( H^{\pm} = \pm 
\frac{1}{2}\left(\frac{1}{r}+\frac{C \cos \phi}{1+Cr \cos \phi}\right) 
\) is the local membrane curvature, and \( dA = d\phi(1+Cr\cos\phi) \).
For 
simplicity, we shall assume here that $\kappa'=\kappa$.  In the full 
model we allow the possibility that the presence of the polymer 
affects the local bending rigidity (i.e. $\kappa'\neq\kappa$).

For our geometry the total mean curvature obeys $\int dA 
H^{\pm}(C,\phi)=\pm 2\pi l$, independent of $C$.  Thus the cost of
bending the 
cylindrical membrane, keeping the polymer distribution homogeneous is:

\begin{equation}
	\Delta E_{hom}^{\pm} \equiv E_{hom}^{\pm}(C)-E_{hom}^{\pm}(0) = 
	2\kappa\int dA\left[H^{\pm}(C,\phi)^2-H^{\pm}(0,\phi)^2\right]~.
	\label{eq:delta-ehom}
\end{equation}

\noindent From this it is obvious that there is always an energy cost to
bend 
the tube in this way.  However, the price is independent of the value 
of $H_0$.

Next, we take into account inhomogeneities in the distribution of 
polymer around the tube.  Such inhomogeneities reduce the energy, if 
polymers move to regions of membrane curvature closer to $H_0$ in both 
the outer and inner monolayers.  
Rewriting Eq.\  (\ref{eq:ehom}) with $\rho$ being now a function of 
$\phi$ we get:
\begin{equation}
E_{inhom}^{\pm}(C) = 2\kappa\int
dA\rho^{\pm}_\phi\left[H^{\pm}(C,\phi)-H_0\right]^2 + 2\kappa\int
dA[1-\rho^{\pm}_\phi] \left[H^{\pm}(C,\phi)\right]^2~,
\label{eq:einhom}
\end{equation}
where $\rho^{\pm}_\phi$ are the polymer distributions on the outer and
inner monolayers.

Subtracting Eq.\  (\ref{eq:einhom}) from Eq.\ (\ref{eq:ehom}), and
taking 
into account conservation of polymer, \( \int dA\rho^{\pm}_\phi = \int 
dA\rho_{0} \), we see that the energy gain, $\Delta
E_{inhom}^{\pm}(C,H_0) 
\equiv E_{hom}^{\pm}(C,H_{0})-E_{inhom}^{\pm}(C,H_{0})$, depends
linearly on the 
spontaneous curvature: 
\begin{equation}
\Delta E_{inhom}^{\pm} = 4\kappa H_{0}\int 
dA\left[\rho_{0}-\rho^{\pm}_\phi\right] H^{\pm}(C,\phi)~.
\label{eq:einhom2}
\end{equation}
Thus, for any polymer distribution with
$\int\rho^{\pm}_\phi H(C,\phi)dA < 
\int\rho_{0}H(C,\phi)dA$, the inhomogeneity lowers the energy; i.e.,
$\Delta E_{inhom}^{\pm}$ is positive and can become arbitrarily large
for large 
values of $H_0$. The detailed calculation (see below) shows
that such configurations indeed exist.

As for the entropy of the system, we assume that the dominant 
contribution is the entropy of mixing of the polymers and lipids.  This 
entropy is larger when the distribution of polymers around the 
cylindrical bilayer is homogeneous, favoring a straight tube.  
However, it does not depend on the spontaneous curvature.  Therefore, 
if $H_0$ is large enough, the energy gain due to $\Delta 
E_{inhom}^{\pm}(C,H_0)$ is larger than the free energy cost coming from 
$\Delta E_{hom}^{\pm}$ and the entropy of mixing.  In this case, the
tube is 
bent at equilibrium.  It remains to be shown that such an equilibrium 
state can occur for reasonable and physical values of the model 
parameters. For this purpose we now turn to the full calculation.


\subsection{Mean Field Calculation}

We neglect fluctuations of the central line curvature $C$, and 
correlations between different segments of the tube.  A section of 
tube of radius $r$ and length $l$, with circular cross section having 
a fixed central line curvature, $C$, has an energy $E = 
E^{+}+E^{-}$, where $E^{+}$ and $E^{-}$ denote the energy of the outer 
and inner leaflets, respectively. $E^{\pm}$ take the form:

\begin{equation}
	E^{\pm} = 2a^{2}\sum_{i,j}
	\left[\kappa(1-\sigma_{ij}^{\pm})+\kappa'\sigma_{ij}^{\pm}\right]
	\left(\frac{1}{r}G^{\pm}(\phi_i) -
	\sigma_{ij}^{\pm}H_{0}\right)^{2},
	\label{eq:micro-en}
\end{equation}

\noindent where \( i = 0, \ldots, \frac{2\pi r}{a}-1 \) is the index 
around the tube and \( j = 0,\ldots,\frac{l}{a}(1+Cr\cos \phi_i) \) is 
the index along the tube (Fig.\  \ref{fig:tube-vars}).  \( 
\sigma_{ij}^{\pm} = 0,1 \) are the occupation variables of the polymer 
on the outer and inner monolayers respectively, and \( G^{\pm}(\phi_i) 
= \pm \frac{1}{2}\left(1+\frac{Cr \cos \phi_i}{1+Cr \cos \phi_i}\right) 
\), i.e. $G^{\pm}(\phi_i)/r$ is the local mean curvature of site 
($ij$).  \( \phi_i = \frac{a}{r}i \) is the angle around the tube as 
shown in Fig.\  \ref{fig:tube-vars}.  By this convention the curvature 
of a site on the outer leaflet has the same magnitude and opposite 
sign of the corresponding site on the inner leaflet.

To find the partition function, as a function of the average polymer 
occupation, we sum over the polymer degrees of freedom:

\begin{equation}
	\Xi^{\pm} = \sum_{ \{ \sigma_{ij}^{\pm} = 0,1 \} } e^{-\beta E^{\pm}
	+ \beta 
	\sum_{i,j} \mu^{\pm} \sigma_{ij}^{\pm}}~.
	\label{eq:partition-fcn}
\end{equation}

The second term in the exponent is a Lagrange multiplier that allows 
us to set the average concentrations on the membrane, \( \rho_{\pm} = 
\sum_{i, j}\sigma_{ij}^{\pm}/N \), to the desired value by 
adjusting the chemical potentials $\mu^{\pm}$. $N$ is the total 
number of sites per monolayer. Note that $\rho_{\pm}$ is the average of
$\rho^\pm_\phi$. From this we calculate the free 
energy:

\begin{equation}
	F^{\pm}(C,\rho) = -K_{B}T\ln (\Xi^{\pm}) + \mu^{\pm} N\rho_{\pm}~.
	\label{eq:free-en}
\end{equation}

\noindent The summation over the polymer occupation degrees of freedom
can be 
carried out, leading to:

\begin{eqnarray}
	\ln (\Xi^{\pm}) & = & \frac{lr}{a^{2}}\int_{0}^{2\pi} d\phi
	(1+Cr\cos\phi)
	\ln \left[ e^{-4\beta \frac{a^{2}}{r^{2}}\kappa 
	\left(G^{\pm}(\phi)\right)^{2}} +e^{-2\beta a^{2}\kappa 
	\left(\frac{1}{r}G^{\pm}(\phi)-H_{0}\right)^{2}+\beta\mu^{\pm}}
	\right],
	\label{eq:partition-fcn4}
\end{eqnarray}

\noindent where the approximation \( \sum_{i} \approx \frac{r}{a} 
\int_{0}^{2\pi}d\phi \) was used.  This approximation is valid when 
the size of a patch is significantly smaller than the radius of the 
tube.  We estimate that $a/r_{0} \approx 10^{-2}$ in our 
system, allowing us to expand Eq.\  (\ref{eq:partition-fcn4}) in powers 
of $a/r$ to second order, and evaluate the integral in 
(\ref{eq:partition-fcn4}):

\begin{eqnarray}
	\ln(\Xi^{\pm}) & = & \frac{\pi l}{2r}\left[
	\frac{4r^{2}}{a^{2}}\ln(1+y^{\pm}) 
	- \frac{2\beta (\kappa + \kappa' y^{\pm} +2\beta 
	a_{2}\kappa'^{2}H_{0}^{2}y^{\pm})}{(1+y^{\pm})\sqrt{1-(Cr)^{2}}} 
	\right. \nonumber \\  & &\left.
	\pm \frac{8\beta \kappa' r H_{0}y^{\pm}}{1+y^{\pm}}-
	\frac{4\beta^{2}\kappa'^{2}a^{2}H_{0}^{2}{y^{\pm}}^{2}}{(1+y^{\pm})^
	{2}\sqrt{1-(Cr)^{2}}}\right],
	\label{eq:partition-fcn8}
\end{eqnarray}

\noindent where \( y^{\pm} = \exp{(\beta\mu^{\pm}-2\beta\kappa'
a^{2}H_{0}^{2})} \). 

We are interested in finding the free energy as function of the
polymer concentration on the membrane, $\rho_{\pm}$, 
rather than the chemical potential $\mu^{\pm}$. Using

\begin{equation}
	\frac{\partial \ln(\Xi^{\pm})}{\partial (\beta\mu^{\pm})} = 
	\frac{2\pi lr}{a^{2}}\rho_{\pm}
	\label{eq:rho1}
\end{equation}

\noindent we get

\begin{equation}
	\rho_{\pm} = 
	\frac{y^{\pm}}{1+y^{\pm}}
	\pm \frac{2\beta\kappa' 
	a^{2}H_{0}^{2}y^{\pm}}{r(1+y^{\pm})^{2}}+O\left( \frac{a^{2}}{r^{2}}
	\right).
	\label{eq:rho2}
\end{equation}

Solving for $y^{\pm}$ gives \( y^{\pm}= \rho_{\pm}(1 \mp 2
\frac{a}{r}\beta\kappa' aH_{0})/(1-\rho_{\pm}) \) to second order in 
$a/r$.  Substituting this into Eq.\  (\ref{eq:free-en}) and 
subtracting the energy for forming the straight tube, \( 
F^{\pm}(0,\rho_{\pm},r) \), we get the free energy cost of bending a
single 
bilayer of radius $r$ within the stack:

\begin{eqnarray}
	F^{\pm}(C,\rho_{\pm},r) - F^{\pm}(0,\rho_{\pm},r) & = & \frac{\pi
	l}{r}\left[\kappa 
	+ \left( \kappa'-\kappa-4\beta\kappa'^{2}a^{2}H_{0}^{2} \right)
	\rho_{\pm}\right.
	\nonumber\\ & & \left.
	+ 4\beta\kappa'^{2}a^{2}H_{0}^{2} \rho_{\pm}^{2}\right]
	\left[ \frac{1}{\sqrt{1-(cr)^{2}}}-1 \right].
	\label{eq:free-en-rho}
\end{eqnarray}

The free energy cost of bending the entire tube is then the 
integral of Eq.\  (\ref{eq:free-en-rho}) over the stack.  In general, 
$\rho_{\pm}$ are functions of $r$.  As we do not know the form of this 
function, we assume for simplicity that the average polymer 
concentration is the same on all the monolayers of the stack, i.e. 
$\rho_{\pm}(r)=\rho_{0}$. Variations in $\rho_{\pm}$ with $r$ do not 
qualitatively change our conclusions. Under this assumption the free 
energy cost of bending the tube is

\begin{eqnarray}
	f(C,\rho_{0}) & = & \frac{1}{d}\int_{0}^{r_{0}} dr
	\left[ 
	F^{+}(C,\rho_0,r) - F^{+}(0,\rho_0,r)+
	F^{-}(C,\rho_0,r) - F^{-}(0,\rho_0,r)
	\right]
	= 
	\nonumber \\
	& = & \frac{2 l \kappa_{tube}(\rho_{0})}{r_{0}^{2}} \ln \left[ 
	\frac{2}{1+\sqrt{1-(Cr_{0})^{2}}} \right],
	\label{eq:integ-free-en}
\end{eqnarray}

\noindent where \( \kappa_{tube} = \frac{\pi r_{0}^{2}}{d}\left[ 
4\beta\kappa'^{2}a^{2}H_{0}^{2}\rho_{0}^{2} + 
(\kappa'-\kappa-4\beta\kappa'^{2}a^{2}H_{0}^{2})\rho_{0} + \kappa 
\right] \), and $d$ is the bilayer spacing in the stack.


The logarithm in Eq.\  (\ref{eq:integ-free-en}) is an increasing 
function of $C$.  Thus the behavior of the tube is dictated by the 
sign of the effective bending modulus of the tube, $\kappa_{tube}$.  
If it is positive, the minimum of $f(C,\rho_{0})$ is at $C=0$, and 
tubes are straight on average.  If, on the other hand, 
$\kappa_{tube}<0$, tubes form tight coils, since the minimum of the 
free energy is at the maximally possible central line curvature 
$C=1/r_0$.  In this case, the free energy of the tube decreases upon 
bending, in agreement with our qualitative argument (see above).

The typical dependence of $\kappa_{tube}$ 
on $\rho_{0}$ for large enough values of $H_0$ is shown in Fig.\
\ref{fig:keff} as a solid line.  When 
$\rho_{0}<\rho_{1}$, $\kappa_{tube}$ is positive and decreases with 
$\rho_{0}$.  In this regime the tube is predicted to be straight on 
the average, but with enhanced fluctuations due to the smaller bending 
modulus.  Although we assumed that the presence of polymer molecules
increases the 
local bending rigidity of the membrane ($\kappa'>\kappa$), their 
mobility makes it {\em easier} to bend the tube.
    
For $\rho_{1}<\rho_{0}<\rho_{2}$, $\kappa_{tube}$ is negative and the
tubes 
form {\em maximally tight} coiled structures.  $\rho_{1}$ is therefore a 
threshold occupancy above which straight tubes are unstable.  Above 
$\rho_{2}$ straight tubes become stable again. However, this regime is 
probably unreachable in our experiments, since too large a polymer 
concentration destroys the bilayers.

For small values of $H_{0}$, $\kappa_{tube}$ is always positive and the 
model does not predict a coiling instability (dashed line in  Fig.\
\ref{fig:keff}). 
Therefore, we now check whether the experimental values of the various 
parameters correspond to a regime in which a coiling instability is 
predicted.  Using pipette aspiration \cite{Evans:97} we measured the 
bending modulus of a bilayer to be $2\kappa \simeq 20 \pm 5 k_{B}T$.  
We assume that $H_{0} > 10 \mu m^{-1}$ because we have observed 
objects that have radii of curvature of the order of, or smaller than 
optical resolution ($\sim 0.2 \mu m$).  We suppose $\kappa^{'} > 
\kappa$; this is consistent with models of composite membranes 
\cite{Lipowsky:95,Hiergeist:96,Helfrich:94} (although the systems 
these models describe are different from ours), and with experiments
(Evans measured 
$\kappa' \simeq 2\kappa$ for membranes with grafted polymers 
\cite{Evans:97}).  Putting these estimates into Eq.\  
(\ref{eq:integ-free-en}) we see that even for small amounts of polymer 
$\kappa_{tube}$ can become negative ($\rho_{1} \leq 0.1$), leading to a 
coiling instability.

The model predicts an inhomogeneous polymer concentration around the 
tube, which we now wish to calculate.  In the expression for $\Xi$ one 
can use a $\phi$-dependent chemical potential $\mu_{\phi}^{\pm}$.  The 
distribution of polymer around the bent tube is then calculated as 
follows:

\begin{equation}
	\rho_{\phi}^{\pm} = \frac{d \ln(\Xi^{\pm}(\{\mu_{\phi}^{\pm}\})}{d
	(\beta\mu_{\phi}^{\pm})} = 
	\frac{1}{1+e^{-2\beta \frac{a^{2}}{r^{2}}\kappa
	G^{\pm}(\phi)^2+2\beta 
	a^{2}\kappa'(\frac{1}{r}G^{\pm}(\phi) - 
	H_{0})^{2}-\beta\mu^{\pm}}}.
	\label{eq:rhoi1}
\end{equation}

The chemical potentials, $\mu^{\pm}$, corresponding to a particular 
total concentration, $\rho_{\pm}$, were found numerically by plotting 
the integral of Eq.\  (\ref{eq:rhoi1}) over $\phi$, as a function of 
$\mu^{\pm}$.  The value of $\mu^{\pm}$ corresponding to the desired 
$\rho_{\pm}$ was then read from the graph.  Using it in Eq.\  
(\ref{eq:rhoi1}), we calculated $\rho_{\phi}^{\pm}$.  Examples of 
such distributions are shown in Fig.\  \ref{fig:rho-of-phi}.

\section{Imaging Model}
\label{image-model}

The model of the previous section predicts maximally curved coils, 
i.e. $C=1/r_{0}$, for $\rho_{1}<\rho_{0}<\rho_{2}$.  In order 
to test this prediction we analyzed the experimental pictures in 
detail.  The images obtained from the microscope are two-dimensional 
projections of the viewed object, and include contributions from 
regions which are out of focus.  Phase contrast microscopy complicates 
the interpretation of the images further, since the intensity at a 
particular point is not a monotonic function of the amount of material 
that the rays of light traverse.  This can create apparent defects, as 
illustrated in Fig.\  \ref{fig:reversal}, and makes resolving images
such 
as Fig.\  \ref{fig:kink}e very difficult.

In order to test the theoretical predictions, we have used numerical 
simulations of various coils to derive virtual projections, and 
compared them with the experimental images.  The simplest objects we 
considered are ideal single and double helices.  We have calculated a 
geometrical phase diagram, for the different possible types of single 
and double helices, their curvatures and lines of contact, as 
described in the Appendix.  Finally, we have considered complex coils 
reminiscent of the complex structures seen in many of our experiments.

By generating simulated curves we can see the effects of various 
imaging parameters on the resulting image.  For example, when the coil 
under observation crosses the focal plane, its shape seems to change 
from a symmetrical arrangement to a series of parallel streaks.  This 
is shown in Fig.\  \ref{fig:tilt}, where an experimental image is 
compared with a simulated ideal helix rotated by ten degrees with 
respect to the imaging plane.  In another case, the image gives the 
impression of a helicity reversal as illustrated in Fig.\  
\ref{fig:reversal}, even though the helix is ideal.  A similar effect 
is seen when a myelin figure is slightly curved. This demonstrates that
even slices of very simple objects may display complex features.
Therefore, it is extremely difficult to reconstruct the objects which
correspond to the experimental images.

Could the tight complex coils observed experimentally (Fig.\  
\ref{fig:kink}) have maximal central line curvature everywhere?  In view
of the difficulties outlined above, we can only partially answer this
question. In order to fully define a curve in three-dimensional space,
two parameters have to be specified at every point.  One choice for such 
parameters is the curvature, $C$, and the torsion, $\tau$.  Our theory 
predicts that $C =1/r_{0}$ when $\rho_1<\rho<\rho_2$. It does not,
however, specify what $\tau$ should be. We investigated various shapes
with $C\approx 1/r_0$ and varying $\tau$ using a numerical simulation.
The program takes two known functions for $C$ 
and $\tau$, and integrates them to form a three-dimensional curve.  
Concentric cylinders are then drawn around this curve.  The resulting 
object can then be rotated and sliced.  Such a slice represents to a 
certain extent the depth of focus of the microscope.  The resulting 
image is then smoothed using a gaussian filter, in order to remove the 
underlying grid, and compared to an experimental image.  

An example of 
this procedure is illustrated in Fig.\  \ref{fig:exp-vs-theo}. The best
fit that we found to the experimental object shown at the top is
displayed at the bottom. This is a maximally curved double helix with a
complex (though periodic) torsion. It may be possible to generate more
complex maximally curved coils, similar to those observed (see Fig.\
\ref{fig:kink}e), using a non-periodic torsion. However, it is difficult
to simulate such objects because of the difficulty in enforcing excluded
volume constraints. Thus, we can only simulate the simplest structures
observed experimentally. The curvatures of these structures are
consistent with the prediction of the model; i.e., they are maximally
curved.

\section{Discussion}
While our model explains the observed phenomena, other models were 
proposed in the literature to explain coiling in other systems.  The 
discussion is devoted to these other possibilities.

The coiling of DMPC-cardiolipin myelin figures was attributed to 
$Ca^{2+}$-mediated membrane-membrane binding \cite{Lin:82}.  Surface 
adhesion was also suggested as the dominant mechanism responsible for 
the coiling of egg-yolk lecithin myelin figures \cite{Mishima:92}.  It 
was proposed that the gain in energy coming from surface adhesion 
overcomes the elastic energy required to bend the tube.  Furthermore 
it was postulated that by twisting back along its length, a rod 
increases the length of the line of contact, thereby increasing the 
area of contact.  An analysis of simple helices shows that this is not 
generally true.  This is shown in the Appendix.  First, there may be 
up to four lines of contact, depending on the type of 
double helix.  Moreover, the length of the line of contact is not 
simply related to how tight the structure is.  In fact the maximum
length of 
the line of contact occurs when the two helices revolve around a large 
radius, thereby being effectively straight.  In our system we 
have found strong evidence that surface adhesion does not play a 
dominant role in determining the morphology.

A generalization of our model may be relevant to other multicomponent
systems.  For example, 
calcium is known to induce a lateral phase separation in a lipid 
mixture containing cardiolipin, thereby creating an inhomogeneous 
distribution.  Egg-yolk lecithin is also a mixture of lipids with the
same 
head-group, and a variety of tail-lengths and degrees of saturation 
\cite{Tattrie:68}.  

Another type of curvature model - the Area Difference 
Elasticity (ADE) model - was discussed in the literature \cite{Miao:94}.
In this approach it is assumed that 
each monolayer has a certain preferred area.  If there is an area 
difference between the monolayers comprising a bilayer, then the 
bilayer will bend in order to accommodate it.  In effect ADE and SC 
both predict the same equilibrium shapes for hollow vesicles.  However, 
ADE cannot explain the coiling of myelin figures.  This is because 
bending a tube, so that it is part of a torus, does not change its 
area.  This is true for each monolayer in the stack, and in particular, 
the area difference between layers is unchanged.

We believe coiling occurs mainly due to the spontaneous curvature 
induced by the presence of the polymer molecules as well as their 
mobility.  Our model emphasizes these aspects, and neglects others 
such as membrane distortions due to the polymer backbone and 
interactions between polymers.  We do not expect these effects to 
change the qualitative behavior of the system.

\section{Acknowledgements}
We thank especially L. Jullien, for his help and encouragement.
We also acknowledge useful exchanges with R. Lipowsky, E. Moses and S. 
Safran. This research was supported by The Israel Science Foundation 
administered by the Israel Academy of Sciences and Humanities - 
Recanati and IDB Group Foundation and The Minerva Foundation.

\appendix
\section{Geometric phase diagram}
\label{phase-diagram}

The requirement that a physical coil does not intersect itself imposes 
a constraint on the relation between the pitch and the radius of a 
helix.  This constraint, as well as the lines of contact, are now 
calculated for single and double helices. Some of the helices studied
here were considered also in \cite{Maritan:00,Stasiak:00}.

\subsection{Single Helix}

The central line of a helical tube can be parameterized as 
follows:

\begin{equation}
	\vec{R}_{1}(t) = (r_{1} \cos(t), r_{1} \sin(t), ht)
	\label{eq:helix}
\end{equation}

\noindent where $r_{1}$ is the radius of the helix around the z-axis, 
$2\pi h$ 
is the pitch, and $t$ is a parameter running along 
the curve.  From this we see that the curvature is \( C = 
r_{1}/(r_{1}^{2}+h^{2}) \), while the torsion is given by \( \tau
= h/(r_{1}^{2}+h^{2}) \).

A myelin figure is then represented by drawing a stack of concentric
tubes having 
a circular cross section around this line:

\begin{equation}
	\vec{R}(t,\theta) = \vec{R}_{1}(t)+r\left[\vec{n}(t)\cos(\theta)+
	\vec{b}(t)\cos(\theta)\right]
	\label{eq:helical-tube}
\end{equation}

\noindent where $\vec{n}(t)$ and $\vec{b}(t)$ are the normal and 
binormal to the curve at point $t$ and $0 < r \leq r_{0}$ as shown in 
Fig.\ \ref{fig:tube-vars}.
The demand that a physical tube does not intersect itself yields two
conditions:

1) The curvature of the central line, $C \leq 1/r_{0}$.  At 
$C= 1/r_{0}$ two consecutive segments along the tube come into 
contact.  From this condition we find that \( h \geq 
\sqrt{r_{1}(r_{0}-r_{1})} \).

2) The pitch of the coil must be such that a segment of the coil does 
not intersect any previous segment.  This requirement can be checked 
numerically by determining the distance between any two points along 
the curve, and demanding this distance to be greater or equal to 
$2r_{0}$.  The result is another constraint on $h(r_{0},r_{1})$.  
Helices for which this distance equals $2r_{0}$ have a contact line 
which is a helix with the same radius and pitch as the central line, 
but shifted by $r_{0}$ along the line of symmetry.
The allowed parameters are summarized in Fig.\ \ref{fig:single-phase}.

\subsection{Double Helices}
\label{double-phase-diagram}

We consider a double helix as two single helices, having radii $r_{1}$ 
and $r_{2}$, that have at least one continuous line of contact.  If 
there is no such contact then they are simply two unconnected single 
helices.  For simplicity we assume that both tubes have the same 
radius $r_{0}$ around their respective central lines, and that these 
central lines wind around the z-axis.  The line of contact is itself a 
helix with the same pitch, and with a radius \( r_2\leq r_{con}\leq
r_1\).  The constraint of having a distance of 
$2r_{0}$ between the two central lines allows us to find $r_{1}$ as a 
function of $r_{2}$ and h.

There are two types of double helices.  Type I double helices have 
\( h > \sqrt{r_{2}(2r_{0}-r_{2})} \). In this case, the two helices wind
around 
their line of contact and their radii obey $r_{1} = 2r_{0}-r_{2}$.  
Double helices of type II obey the condition \( h <
\sqrt{r_{2}(2r_{0}-r_{2})} \). In this case, the coils are 
intermingled and wind around a hole in the middle. Type II
double helices have at least two lines of contact (an example is shown
in structure d of Fig.\ \ref{fig:sample-helices}. This is summarized as
a phase diagram in Fig.\  \ref{fig:dbl-phase}.


We now find the lines of contact between the two coils of a double 
helix.  Using Eq.\  (\ref{eq:helix}) we define the central lines of the 
two strands:

\begin{eqnarray}
	\vec{R}_{1}(t_{1}) & = & (r_{1}\cos(t_{1}), r_{1}\sin(t_{1}),
	ht_{1})
	\nonumber \\
	\vec{R}_{2}(t_{2}) & = & (r_{2}\cos(t_{2}+\eta),
	r_{2}\sin(t_{2}+\eta),
	ht_{2})~.
	\label{eq:dbl-helice}
\end{eqnarray}

\noindent The phase $\eta$ is $\eta=\pi$ for double helices of types I
and II, and for loose helices $\eta$ is determined by the requirement
that there exists at least one line of contact. The line of contact is
then given by \( \vec{D}(t_{1},t_{2}) 
= \vec{R}_{1}(t_{1}) - \vec{R}_{2}(t_{2}) \).  From the requirement 
that $\vec{D} \perp \partial \vec{R}_{i}/\partial t_{i}$ and 
that $|\vec{D}| = 2r_{0}$ we find that the line of contact is a helix 
with the same pitch as the other two helices, and a radius:

\begin{equation}
	r_{c} =
	\frac{1}{2}\sqrt{r_{1}^{2}+r_{2}^{2}-2r_{1}r_{2}\cos(\Delta)}
	\label{eq:r-contact},
\end{equation}

\noindent where $\Delta = \sqrt{b/d - 2 \pm 
2\sqrt{1-b/d+1/d^{2}}}$, $d = 
h^{2}/(r_{1}r_{2})$ and $b = 
(4r_{0}^{2}-r_{1}^{2}-r_{2}^{2})/(r_{1}r_{2})$.

Double helices of type I have one line of contact, while type II 
helices typically have two lines of contact.  At the dotted line in
Fig.\  
\ref{fig:dbl-phase} the two lines merge.  Type II helices 
lying on the line of self intersection also have lines of self contact,
for a total 
of four lines of contact.  The maximal length of contact lines occurs 
when one helix lies within the other and their radii go to infinity.  
In this case the total length of contact is four times the length of 
the central line of each helix. Thus, formation of a tighter helix does
not necessarily mean that the line 
of contact increases, not to mention the area of contact.  The area of 
contact between two cylinders depends on the angle at which they come 
into contact.  Moreover, when there are two lines of contact close 
together (for example near the dotted line in Fig.\
\ref{fig:dbl-phase})
the areas 
of contact must overlap to some degree.


\newpage

\begin{figure}[h]
 	\epsfxsize=75mm
 	\vspace{0.5cm}
	\caption{Image of a myelin figure at very low polymer 
	concentration.   The scalebar represents 20 $\mu$m.}
\label{fig:straight}
\end{figure}

\begin{figure}[h]
	\epsfxsize=75mm
	\vspace{0.5cm}
	\caption{When the instability begins at the tip, the tube loops 
	back upon itself forming a globular structure.  Such a structure 
	is seen here (a) 0 (b) 83 (c) 109 (d) 155 (e) 221 (f) 371 seconds 
	after onset of bending.   The scalebar represents 10 $\mu$m.}
\label{fig:ball}
\end{figure}

\begin{figure}[h]
	\epsfxsize=75mm
	\vspace{0.5cm}
	\caption{A sequence of images depicting a single helix being 
	mechanically stretched, and returning to a maximally curved 
	configuration.  The helix behaves like a spring, responding to the 
	stretching force by elongating uniformly.  When the force is 
	removed the coil retracts.  This behavior suggests a restoring 
	force, rather than surface adhesion.  Times are: (a) before 
	stretching, and (b) 67 (c) 102 (d) 159 (e) 215 (f) 325 (g) 393 
	seconds after initiation of stretching. Snapshots (d-g) were taken
	after the force was removed.  The scalebar represents 10 
	$\mu$m.}
	\label{fig:spring}
\end{figure}

 \begin{figure}[h]
 	\epsfxsize=75mm
 	\vspace{0.5cm}
	\caption{Fluorescence image of a double helix.   The scalebar 
	represents 10  $\mu$m.}
 	\label{fig:dbl-coil}
 \end{figure}

\begin{figure}[h]
	\epsfxsize=75mm
	\vspace{0.5cm}
	\caption{Formation of a complex coil.  The tube becomes unstable 
	locally, forming a hairpin which gradually curls up.  The intervals 
	between snapshots (a-d) are 45 seconds long. Structure (e) was
	observed 13 minutes after onset.  
	 The scalebar represents 10 $\mu$m.}
\label{fig:kink}
\end{figure}

\begin{figure}[h]
	\epsfxsize=75mm
	\vspace{0.5cm}
	\caption{A double helix coming into contact with another myelin
	figure.  
	As the coil grows, it pushes the other figure aside.  Fifteen 
	minutes after initial contact, the two myelin figures are no 
	longer touching.  This is another demonstration that adhesion is 
	not important in this system.  The scalebar represents 10 $\mu$m.}
	\label{fig:contact}
\end{figure}

\begin{figure}[h]
	\epsfxsize=75mm
	\vspace{0.5cm}
	\caption{Fluorescence image of a myelin figure showing that there 
	is polymer inside.  Inset: fluorescence intensity along the 
	section.   The scalebar represents 10 $\mu$m.}
	\label{fig:flo-tube}
\end{figure}

\begin{figure}[h]
	\epsfxsize=75mm
	\vspace{0.5cm}
	\caption{Schematic representation of the heuristic argument.  
	Starting from a straight tube with the same polymer concentration on
	the inner and outer monolayers (a), the cost of bending the tube 
	keeping $\rho(\phi)$ homogeneous (b) is $E_{hom}^{\pm}$.  Allowing 
	the polymer to diffuse around the tube (c) to the configuration
	shown 
	in Fig.\  \ref{fig:rho-of-phi} lowers the energy by $\Delta 
	E_{inhom}^{\pm}$, but also lowers the entropy.  The black (white)
	color in 
	(c) corresponds to areas of small (large) polymer concentration.  
	Both $E_{hom}$ and the entropy are independent of 
	$H_{0}$, while $E_{inhom}^{\pm}$ is linear in $H_{0}$.  Thus for 
	large values of $H_{0}$ it is preferable to bend the tube.}
\label{fig:scheme}
\end{figure}

\begin{figure}[h]
	\epsfxsize=75mm
	\vspace{0.5cm}
	\caption{A small section of a coiled stack of concentric membranes 
	showing the directions of the the normal, $\vec{n}$, and the 
	binormal, $\vec{b}$.  Each monolayer, having a radius $r$ is 
	divided into patches with running indices $i$ and $j$.}
\label{fig:tube-vars}
\end{figure}

\begin{figure}[h]
	\epsfxsize=75mm
	\vspace{0.5cm}
\caption{The effective bending modulus of the tube, $\kappa_{tube}$, is
parabolic in the average occupancy, $\rho$. We have used the following
values of the parameters: $\kappa=10 k_B T$, $\kappa'=2\kappa$ and
$r_0=5\mu$m. We find that $\kappa_{tube}$ depends on $a$ and $H_0$ only
through the product $aH_0$.
The solid curve represents $\kappa_{tube}$ for 
$aH_0=0.3$. When $aH_0$ is large
enough ($aH_0>0.19$ for the values of $\kappa$
and $\kappa'$ we have used), $\kappa_{tube}<0$ between
$\rho_1$ and $\rho_2$. For smaller values of $aH_0$,
$\kappa_{tube}>0$ for all values of $\rho$. The dashed curve
corresponds to $aH_0=0.16$.}
	\label{fig:keff}
\end{figure}

\begin{figure}[h]
	\epsfxsize=75mm
	\vspace{0.5cm}
	\caption{The local polymer concentration as a function of the 
	angle around the tube is shown for two average polymer
	concentrations: $\rho=0.5$ (solid line) and $\rho=0.1$ (dashed 
	line). The following parameters have been used: 
	$r/a=100$, $rC=0.99$, $aH_{0}=0.3$, $\kappa=10k_{B}T$, 
	$\kappa'=2\kappa$, and $\rho=0.5$ (solid line), $\rho=0.1$ (dashed 
	line).}
\label{fig:rho-of-phi}
\end{figure}

\begin{figure}[h]
	\epsfxsize=75mm
	\vspace{0.5cm}
	\caption{Top: Fluorescence image of a double helix showing that 
	the helicity seems to reverse where the coil intersects the focal 
	plane.  This is an artifact of the imaging geometry, as can be 
	seen for the theoretical {\em ideal} double helix tilted at
	$15^{\circ}$ at the bottom.   The scalebar represents 10 $\mu$m.}
	\label{fig:reversal}
\end{figure}

\begin{figure}[h]
	\epsfxsize=75mm
	\vspace{0.5cm}
	\caption{A comparison between an experimental helix (left) with a 
	model ideal double helix (right).  This image shows the effect of 
	tilting the coil by 10$^{\circ}$ with respect to the imaging plane.  
	Notice the cut has a symmetric appearance at the bottom, where the
	imaging plane 
	slices through the center of the coil, shifting gradually towards 
	an array of parallel smudges at the top. The scalebar represents 10
	$\mu$m.}
\label{fig:tilt}
\end{figure}

\begin{figure}[h]
	\epsfxsize=75mm
	\vspace{0.5cm}
	\caption{Top: experimental image of a double helix.  Bottom: 
	cross section of a theoretical double helix having maximal 
	curvature ($C=1/r_0$), and sinusoidal torsion,
	$\tau=2+2\sin\left\{\frac{2\pi}{31} \left[ i+1.8\sin(\frac{2\pi 
	i}{31}+\frac{3\pi}{2})\right]\right\}$. The central line of the
	image is 
	shifted by $r_{0}$ below the imaging plane. The scalebar represents
	10 $\mu$m.}
	\label{fig:exp-vs-theo}
\end{figure}

\begin{figure}[h]
	\epsfxsize=75mm
	\vspace{0.5cm}
	\caption{The line of self-contact for single-helices.  $r_0$ is the
	outer radius of the tube. $2\pi h$ and $r_1$ are the pitch and the
	radius of the central line of the helix, respectively. The
	requirement 
	that a physical coil does not intersect itself gives two 
	conditions. The dashed line corresponds to the first condition, 
	while the solid lines corresponds to the second condition, as 
	explained in the text.}
	\label{fig:single-phase}
\end{figure}

\begin{figure}[h]
	\epsfxsize=75mm
	\vspace{0.5cm}
	\caption{Geometrical phase diagram for double helices.  
	$r_{0}$ is the outer radius of the tubes and $2\pi h$ is the pitch
	of the central line of the two helices. $r_1$ and $r_{2}$ are the
	radii of 
	the central lines of the two helices. Our convention is that $r_{2}
	\leq r_{1}$. 
	There are two types of geometries, separated by the dashed line in
	the figure.  Type I: two helices winding 
	around each other and type II: interlaced helices revolving around 
	a central hole.  The lower bound is the limit below which a helix 
	intersects itself.  The top-right bound is the line of symmetry, 
	where both helices have the same radius.}
	\label{fig:dbl-phase}
\end{figure}

\begin{figure}[h]
	\epsfxsize=75mm
	\vspace{0.5cm}
	\caption{Simulations of representative ideal theoretical helices.
	Letters 
	are keyed to the diagram in Fig.\ \ref{fig:dbl-phase}.  Most of the 
	double helices observed experimentaly resemble type (c), although
	some of them are more symmetric than the example shown.}
	\label{fig:sample-helices}
\end{figure}

\end{document}